\documentclass[11pt,twoside,twocolumn]{article}

\usepackage{amsfonts,amsmath,amssymb,bm,abstract,balance,natbib,graphicx}

\usepackage[active]{srcltx}

\title{\textbf{Effective attraction between oscillating electrons \\
in a plasmoid via acoustic waves exchange}}

\author{Maxim Dvornikov$^{a,b}$
\\
$^{a}$\small{\ N.~V.~Pushkov Institute of Terrestrial Magnetism, Ionosphere} \\
\small{and Radiowave Propagation (IZMIRAN),} \\
\small{142190, Troitsk, Moscow Region, Russia;} \\
$^{b}$\small{\ Institute of Physics, University of S\~{a}o Paulo,} \\
\small{CP 66318, CEP 05315-970 S\~{a}o Paulo, SP, Brazil} \\
\small{E-mail: maxdvo@izmiran.ru}}

\date{}

\begin{document}

\twocolumn[\maketitle
\begin{onecolabstract}
We consider the effective interaction between electrons due to the
exchange of virtual acoustic waves in a low temperature plasma.
Electrons are supposed to participate in rapid radial oscillations
forming a spherically symmetric plasma structure. We show that
under certain conditions this effective interaction can result in
the attraction between oscillating electrons and can be important
for the dynamics of a plasmoid. Some possible applications of the
obtained results to the theory of natural long-lived plasma
structures are also discussed.
\\

PACS numbers: 52.35.Dm, 52.35.Sb, 92.60.Pw
\\
\end{onecolabstract}]


\section{Introduction}

The interaction between a test charged particle, embedded in a
warm plasma, and the collective response of a plasma results in
the concept of a dressed particle. Accounting for the collective
plasma interaction, for a particle at rest one gets the screening
of the vacuum Coulomb interaction leading to the Debye-H\"{u}ckel
potential~\citep{LanLif80}. \citet{NamAka85} showed that a test
particle moving in plasma is a source of ion acoustic waves
forming in its wake. If the speed of a test particle is close to
the speed of the ion acoustic waves propagation, the resulting
wake potential prevails the Debye-H\"{u}ckel interaction leading
to the effective attraction of charged particles of the same
polarity~\citep{NamAka85}. This process may be responsible for the
formation of complex structures in dusty
plasmas~\citep{ShuMam02,Tsy07}.

There is, however, another possibility for charged particles of
equal polarities to experience an effective attraction in a
plasma. It happens when charged particles interact via an acoustic
wave appearing in the neutral component of
plasma~\citep{VlaYak78}. This process is analogous to the phonon
exchange between two electrons in metal~\citep{Mad80}. For the
implementation of this mechanism the temperature of plasma should
be at least less than the ionization potential of an atom --
otherwise neutral atoms can exist only in a small fraction. The
results of~\citet{VlaYak78} may have various applications. In
particular, they can be used for the explanation of the stability
of atmospheric plasma structures~\citep{Ste99}.

In the present work, using the formalism of~\citet{VlaYak78}, we
will study the effective attraction between charged particles
participating in spherically symmetric oscillations of a low
temperature plasma. Recently radial plasma pulsations were
considered by~\citet{DvoDvo07,Dvo10,Dvo11}, in frames of both
quantum and classical approaches, as a theoretical model of
natural plasmoids~\citep{Ste99}. Unlike~\citet{VlaYak78}, in the
present work we will suggest that the stability of a plasma
structure is provided by some other mechanisms, like various
nonlinear~\citep{SkoHaa80,LaeSpa84,Dvo11}, or
quantum~\citep{HaaShu09} effects.

This work is organized as follows. In Sec.~\ref{EIG} we develop
the general theory of the effective interaction between two
charged particles via the exchange of an acoustic wave and apply
it to electrons performing spherically symmetric oscillations in
plasma. Then, in Sec.~\ref{PAF}, we evaluate the parameters of the
effective potential and analyze the conditions when it results in
the attraction between oscillating electrons. In Sec.~\ref{APPL}
we consider the application of the described effective interaction
to the dynamics of natural long-lived plasmoids. Finally, in
Sec.~\ref{CONCL}, we summarize our results.

\section{Effective interaction in a spherically symmetric
plasma structure\label{EIG}}

In this section we formulate the general dynamics of a spherically
symmetric oscillation of a low temperature plasma. The electron
temperature should be less than the ionization potential of an
atom. In this case both electron, ion, and neutral components can
coexist in a plasma. For hydrogen plasma this critical temperature
is $\sim 10^5\thinspace\text{K}$. Then we consider the effective
interaction between two oscillating electrons due to the exchange
of a virtual acoustic wave and establish conditions when this
interaction corresponds to a repulsion and an attraction between
particles.

Suppose that one has excited a plasma oscillation in a low
temperature plasma. In this situation both electrons and ions will
oscillate with respect to neutral atoms. Note that electrons will
oscillate with higher frequency than that of ions since their
mobility is much bigger. If the density of neutral atoms is
sufficiently high, charged particles will collide with neutral
atoms generating perturbations of their density or even waves in
the neutral component of plasma. Typically this process will
result in the acoustic waves emission and finally in the energy
dissipation in the system. However, if an acoustic wave is emitted
coherently to be absorbed by another charged particle, one can
expect the appearance of an effective interaction. The analogous
phenomenon is well known in solid state physics~\citep{Mad80}. It
was suggested by~\citet{VlaYak78} that such an effective
interaction may well happen in a low temperature plasma.

In this work we will mainly discuss spherically symmetric plasma
oscillations. The possibility of the existence of such a plasma
structure is predicted in both
classical~\citep{SkoHaa80,LaeSpa84,Dvo11} and
quantum~\citep{DvoDvo07,HaaShu09} cases. Stable classical
plasmoids can exist due to the various plasma nonlinearities,
which arrest the collapse of oscillations~\citep{Gol84}, whereas
the stability of quantum plasmoids is provided by the additional
quantum pressure. In our study we do not specify physical
processes underlying the stability of the plasma structure in
question.

Omitting the motion of ions we can represent the plasma
characteristics, like the electron density $n_e$, the electron
velocity $\mathbf{v}_e$, and the electric field $\mathbf{E}$, in
the following way:
\begin{align}\label{eloscdef}
  n_e = & n_0 + n_1 \mathrm{e}^{- \mathrm{i} \omega t} +
  n_1^{*{}} \mathrm{e}^{\mathrm{i} \omega t} + \dotsb,
  \notag
  \\
  \mathbf{v}_e = & \mathbf{v}_1 \mathrm{e}^{- \mathrm{i} \omega t} +
  \mathbf{v}_1^{*{}} \mathrm{e}^{\mathrm{i} \omega t} + \dotsb,
  \notag
  \\
  \mathbf{E} = & \mathbf{E}_1 \mathrm{e}^{- \mathrm{i} \omega t} +
  \mathbf{E}_1^{*{}} \mathrm{e}^{\mathrm{i} \omega t} + \dotsb,
\end{align}
where $n_0$ is the background electron density and the index ``1''
stays for the oscillating parts of the electron density $n_1$, the
electron velocity $\mathbf{v}_1$, and the electric field
$\mathbf{E}_1$. As we study spherically symmetric plasma
oscillation all the functions in Eq.~\eqref{eloscdef} depend only
on the radial coordinate $r$ as well as the vectors $\mathbf{v}_1$
and $\mathbf{E}_1$ have only the radial component.

Note that the frequency of the plasma oscillation $\omega$ in
Eq.~\eqref{eloscdef} is typically less than the Langmuir frequency
for electrons $\omega_p = \sqrt{4 \pi e^2 n_0/m}$, where $e>0$ is
the proton charge and $m$ is the mass of the electron. The
discrepancy between $\omega$ and $\omega_p$ depends on the
physical processes proving the plasma structure stability. For
example, for a classical plasmoid, which is treated
by~\citet{SkoHaa80,LaeSpa84,Dvo11} with help of the perturbation
theory, it can be about 10\%. However, if a system possesses a
high degree of nonlinearity, the difference between $\omega$ and
$\omega_p$ can be significant. We will suppose that $\omega = \xi
\omega_p$, with $\xi < 1$.

\begin{figure}
  \centering
  \includegraphics[scale=.4]{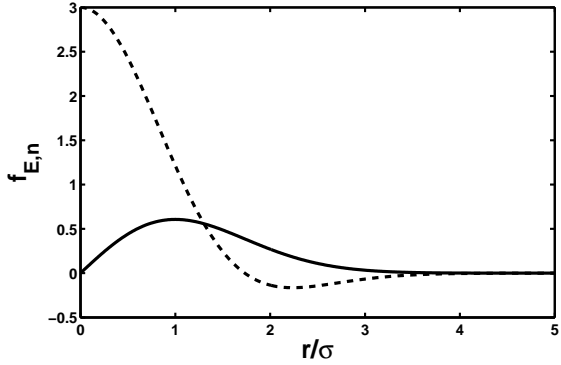}
  \caption{\label{E1n1}
  The functions $f_E = E_1/A\sigma$ (solid line)
  and $f_n = - 4 \pi e n_1 / A$
  (dashed line) versus the radius of a plasmoid.}
\end{figure}

The quantities $n_1$, $\mathbf{v}_1$, and $\mathbf{E}_1$ are
related by the plasma hydrodynamic equations derived
by~\citet{SkoHaa80},
\begin{equation}\label{vEn}
  \mathbf{v}_1 = - \frac{\mathrm{i}e}{m\omega_p} \mathbf{E}_1,
  \quad
  (\nabla \cdot \mathbf{E}_1) = - 4 \pi e n_1.
\end{equation}
Therefore we can describe the plasma oscillation in terms of only
one function, e.g., $E_1 = |\mathbf{E}_1|$. The form of this
function depends on the peculiar type of a plasmoid and it is
rather difficult to find it analytically. Thus one should rely
only on the numerical simulations of the plasmoid dynamics.
Nevertheless we can use the following \textit{ansatz} proposed
by~\citet{And83},
\begin{equation}\label{Edistr}
  E_1(r) = A r \exp
  \left(
    -\frac{r^2}{2\sigma^2}
  \right),
\end{equation}
which describes quite accurately the dynamics of a
plasmoid~\citep{Ban02}. Now a plasmoid is described in terms of
the amplitude $A$ and the width $\sigma$. In Fig.~\ref{E1n1}
we show the schematic behaviour of the perturbations of the
electric field and the normalized number density of electrons $f_n
= (3-r^2/\sigma^2) \exp(-r^2/2\sigma^2)$.

\begin{figure}
  \centering
  \includegraphics[scale=1]{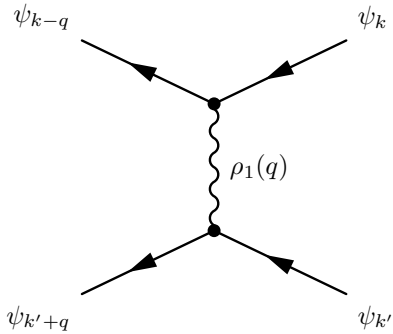}
  \caption{\label{feyndiageescatt}
  The schematic illustration of the exchange between two oscillating
  electrons by a virtual acoustic wave. Initially electrons are in the
  states $\psi_k$ and $\psi_{k'}$ whereas their final states are
  $\psi_{k-q}$ and $\psi_{k'+q}$. The definition of the ``quantum'' of
  an acoustic wave, $\rho_1$, is given in Eq.~\eqref{ngprogosc}.}
\end{figure}

As we mentioned above, rapidly oscillating electrons, described by
Eqs.~\eqref{eloscdef} and~\eqref{Edistr}, can interact with
neutral atoms generating oscillations in the neutral component of
plasma. Under certain conditions these acoustic waves can be
absorbed by other electrons producing an effective interaction
between charged particles. The process of the exchange of a
virtual acoustic wave between two electrons is schematically shown
in Fig.~\ref{feyndiageescatt}.

At the second order of the perturbation theory the matrix element
of the electrons interaction due to the exchange of the
``quantum'' of an acoustic wave has the form~\citep{Mad80},
\begin{equation}\label{2ordme}
  V_{k k' q} =
  \frac{2 \hbar \omega_q |M_{k,k-q}|^2}
  {[E(k)-E(k-q)]^2 - (\hbar \omega_q)^2},
\end{equation}
where $\omega_q$ is the frequency of a virtual acoustic wave,
$E(k)$ are the energy levels of an electron participating in rapid
oscillations, and $M_{k,k-q}$ is the matrix element of the
electron's interaction with the ``quantum'' of an acoustic wave
taken in the first order of the perturbation theory. Note that an
electron state in Eq.~\eqref{2ordme} is specified with help of the
single quantum number $k$ rather than of a vector $\mathbf{k}$
since here we consider a radial motion of charged particles.

To determine the type of the effective interaction~\eqref{2ordme}
one should examine the sign of the denominator in
Eq.~\eqref{2ordme}. The situation when $|E(k)-E(k-q)| > \hbar
\omega_q$ corresponds to the repulsion between electrons and
$|E(k)-E(k-q)| < \hbar \omega_q$ to the effective attraction. In
the next section we will evaluate the energy change $\Delta E =
|E(k)-E(k-q)|$ in a collision with a neutral atom as well as the
typical frequency of an acoustic wave $\omega_q$.

\section{Parameters of the effective potential\label{PAF}}

In this section we evaluate the parameters of the effective
interaction~\eqref{2ordme} and show that under certain conditions
it can be attractive. Then we estimate the magnitude of the
effective interaction and compare it with the typical kinetic
energy of oscillating electrons.

To evaluate the energy change $\Delta E$ of an electron while it
emits an acoustic wave we should notice that this acoustic wave is
generated when an electron collides with neutral atoms. Thus
knowing the total power of acoustic waves emitted and the mean
number of collisions with neutral atoms per unit time we can
evaluate the energy change in a collision.


We describe the dynamics of neutral gas using a set of the
inhomogeneous hydrodynamic equations proposed by~\citet{Ing66},
\begin{align}\label{hydrng}
  \frac{\partial \rho_a}{\partial t} & + \nabla \cdot (\rho_a \mathbf{u}) = 0,
  \qquad
  \notag
  \\
  \rho_a \frac{\partial \mathbf{u}}{\partial t} & + (\mathbf{u}\nabla)\mathbf{u} +
  \nabla p = \mathbf{F},
\end{align}
where $\rho_a$ is the mass density of neutral gas, $\mathbf{u}$ is
its velocity, $p$ is the neutral gas pressure and $\mathbf{F}$ is
the rate of the momentum transfer per unit volume from an external
source, which is the electron subsystem in our case, to the
neutral gas. In Eq.~\eqref{hydrng} we suppose that the generation
of acoustic waves happens at the constant entropy. We should also
supply Eq.~\eqref{hydrng} with the equation of state of the
neutral gas, $p = p(\rho_a)$.

Taking into account the oscillatory character of the electron
number density~\eqref{eloscdef} we get that the external force
$\mathbf{F}$ should also have the similar behaviour, $\mathbf{F} =
\mathbf{F}_1 \mathrm{e}^{- \mathrm{i} \omega t} +
\mathbf{F}_1^{*{}} \mathrm{e}^{\mathrm{i} \omega t}$. Linearizing
Eq.~\eqref{hydrng} we obtain that the pressure, the density, and
the velocity of the neutral gas also oscillate on the same
frequency $\omega$:
%
%
\begin{align}\label{ngprogosc}
  p = & p_0 + p_1 \mathrm{e}^{- \mathrm{i} \omega t} +
  p_1^{*{}} \mathrm{e}^{\mathrm{i} \omega t},
  \notag
  \\
  \rho_a = & \rho_0 + \rho_1 \mathrm{e}^{- \mathrm{i} \omega t} +
  \rho_1^{*{}} \mathrm{e}^{\mathrm{i} \omega t},
  \notag
  \\
  \mathbf{u} = & \mathbf{u}_1 \mathrm{e}^{- \mathrm{i} \omega t} +
  \mathbf{u}_1^{*{}} \mathrm{e}^{\mathrm{i} \omega t},
\end{align}
where $p_0$ and $\rho_0$ are the equilibrium values. As in
Eq.~\eqref{eloscdef} the index ``1'' stays here for the perturbed
quantities and higher harmonics are omitted.

Using Eq.~\eqref{ngprogosc} we can represent the linearized
Eq.~\eqref{hydrng} for the neutral gas pressure as the
inhomogeneous Helmholtz equation,
\begin{equation}\label{Hgeqp}
  \nabla^2 p_1 + k^2 p_1 = (\nabla \cdot \mathbf{F}_1),
\end{equation}
where $k = \omega/c_a$ is the wave vector of the acoustic wave and
$c_a = \sqrt{(\partial p/\partial \rho_a)_S}$ is the sound
velocity in the neutral gas, with the derivative being taken at
the constant entropy. Note that acoustic waves will be emitted
with the same frequency as the plasma oscillation, $\omega_q =
\omega$.

Now we should specify the type of the interaction between
electrons and neutral atoms. According to the definition of the
external force it can be represented as $\mathbf{F}_1 = - n_a
\nabla U$, where $n_a = \rho_0/M_a$ is the background number
density of neutral atoms and $M_a$ is the mass of an atom. We also
suggest that only oscillating electrons contribute to the
effective potential $U(\mathbf{r})$ of the interaction of an
acoustic wave with electrons,
\begin{equation}\label{Upot}
  U(\mathbf{r}) = \int \mathrm{d}^3\mathbf{r}'
  n_1(\mathbf{r}') K(\mathbf{r}-\mathbf{r}'),
\end{equation}
where $K(\mathbf{r})$ is the potential of the interaction between
an electron and a neutral atom.

Let us choose the charge-dipole potential first proposed
by~\citet{Buc37}, which is a good model of the interaction between
electrons and neutral atoms in a low energy plasma,
\begin{equation}\label{Kenergy}
  K(\mathbf{r}) =
  - \frac{\alpha e^2}{(|\mathbf{r}|^2 + r_0^2)^2},
\end{equation}
where $r_0$ is the cut-off radius, which is of the order of the
atomic size, and $\alpha \sim r_0^3$ is the electric dipole
polarizability of an atom. The corrections to Eq.~\eqref{Kenergy}
due to the collective plasma effects were studied
by~\citet{Red87}. It was found that for a hydrogen plasma the
deviations from Eq.~\eqref{Kenergy} are negligible at the low
plasma temperature $\sim 10^3\thinspace\text{K}$ used below in
this section.

Taking into account the general solution of Eq.~\eqref{Hgeqp},
\begin{equation}
  p_1(\mathbf{r}) = -\frac{1}{4\pi} \int \mathrm{d}^3\mathbf{r}'
  \frac{\mathrm{e}^{\mathrm{i}k|\mathbf{r}-\mathbf{r}'|}}{|\mathbf{r}-\mathbf{r}'|}
  (\nabla \cdot \mathbf{F}_1),
\end{equation}
we can represent the total power, radiated in the form of the
acoustic waves, in the following form:
\begin{align}\label{toten}
  \langle \dot{\mathcal{E}} \rangle = & \int \mathrm{d}^3\mathbf{r}
  (\mathbf{F}_1^{*{}}\mathbf{u}_1 + \mathbf{F}_1 \mathbf{u}_1^{*{}})
  \notag
  \\
  & =
  \frac{8 \pi n_a k^3}{\omega M_a}
  \left(
    \int_0^\infty \mathrm{d}r \ r U(r) \sin kr
  \right)^2.
\end{align}
To derive Eq.~\eqref{toten} we suppose that the function $U(r)$
rapidly decreases at large distances, cf. Eq.~\eqref{Edistr}.

Now let us evaluate the number of collisions with neutral atoms
per unit time at rapid oscillations of electrons. For this purpose
it is more convenient to use the Lagrange variables: $(r,t) \to
(\rho,\tau)$. Suppose that an oscillating electron is at the
distance $\rho$ from the center of the system and its law of
motion has the form, $r = \rho + A_0(\rho)\sin \omega \tau$, where
$A_0$ is the amplitude of oscillations. Taking into account the
continuity equation in Lagrange variables for a spherically
symmetric system found by~\citet{Dvo11}, $n_0 \rho^2 = n_e r^2
(\partial r/\partial \rho)$, and supposing that the amplitude
$A_0$ is not so large, i.e. $n_1(r) \approx n_1(\rho)$, we obtain
the number of collisions per unit time as
\begin{align}\label{dotN}
  \dot{\mathcal{N}} = & 4 n_a \sigma_s \omega
  \int_0^\infty A_0(\rho) n_1(\rho) \rho^2 \mathrm{d}\rho
  \notag
  \\
  & =
  4 \sigma_s \omega \frac{n_a}{n_0}
  \int_0^\infty n_1(\rho) \mathrm{d}\rho
  \notag
  \\
  & \times
  \int_0^\rho r^2 n_1(r) \mathrm{d}r,
\end{align}
where $\sigma_s$ is the cross section of the electron scattering
off a neutral atom. To derive Eq.~\eqref{dotN} we suppose that
$A_0(0) = 0$, i.e. oscillations vanish at the center of the
system.

Now we can evaluate the energy change in a collision with a
neutral atom as $\Delta E \sim \langle \dot{\mathcal{E}} \rangle /
\dot{\mathcal{N}}$. As it was mentioned above, the situation when
$|\Delta E|$ is less than the typical energy of the ``quantum'' of
an acoustic wave, $\hbar \omega_q$, corresponds to the effective
attraction between electrons. Besides this condition, for the
effective interaction to become important for the dynamics of the
system, its magnitude should be comparable with the typical
kinetic energy of oscillating electrons. That is why we have to
evaluate the matrix element of an acoustic wave generation by an
oscillating electron $M_{k,k-q}$ in Eq.~\eqref{2ordme}.

This matrix element can be expressed in the following form:
\begin{equation}\label{M1partme}
  M_{k,k-q} =
  \int \mathrm{d}^3\mathbf{r}
  \psi_k^{*{}}(r) V(\mathbf{r}) \psi_{k-q}(r),
\end{equation}
where $V(\mathbf{r})$ is the energy of interaction of an electron
and the acoustic field. The wave function of an electron
$\psi_k(r)$ in Eq.~\eqref{M1partme} can be taken in the form of a
normalized spherical wave,
\begin{gather}\label{psidef}
  \psi_k(r) = \sqrt{\frac{k}{\pi}} \frac{\sin(kr)}{r},
  \notag
  \\
  \int \mathrm{d}^3\mathbf{r} \psi_k^{*{}}(r) \psi_{k'}(r) =
  2 \pi k
  \delta(k-k'),
\end{gather}
since we study a localized oscillation. Generally speaking, in a
realistic situation the wave function of an electron can differ
from that in Eq.~\eqref{psidef}. However, to get a rough estimate
we can choose it in such a form.

Analogously to Eq.~\eqref{Upot} we suggest that an electron
scatters off the density perturbation caused by an acoustic wave,
i.e. at the first order of the perturbation theory the effective
potential $V(\mathbf{r})$ has the following form proposed
by~\citet{VlaYak78}:
\begin{equation}\label{Vpot}
  V(\mathbf{r}) =
  \frac{1}{M_a} \int \mathrm{d}^3\mathbf{r}'
  \rho_1(\mathbf{r}')
  K(\mathbf{r}-\mathbf{r}'),
\end{equation}
where the energy of interaction between an electron and a neutral
atom $K(\mathbf{r})$ is given in Eq.~\eqref{Kenergy}. Note that in
Eq.~\eqref{Vpot} we take into account only the oscillating part of
the neutral gas density $\rho_1$, cf. Eq.~\eqref{ngprogosc}.
Accounting for Eqs.~\eqref{Kenergy}
and~\eqref{M1partme}-\eqref{Vpot} we obtain the matrix element
$M_{k,k-q}$ as
\begin{align}\label{Mfinal}
  M_{k,k-q} = &
  \text{sign}(k-q) \sqrt{\left| 1-\frac{q}{k} \right|}
  \notag
  \\
  & \times
  \frac{2 \alpha e^2}{r_0} \frac{k^2 \pi^3 n_a}{c_a^2 M_a}
  \notag
  \\
  & \times
  \mathrm{e}^{-k r_0}
  \int_0^\infty \mathrm{d}r \ r U(r) \sin kr.
\end{align}
%
%
%
Here we take into account that both $\rho_1$ and $\rho_1^{*{}}$
contribute to the matrix element.

It is convenient to compare the potential of the effective
interaction between oscillating electrons~\eqref{2ordme} with the
typical value of the kinetic energy of an oscillating electron.
Using Eq.~\eqref{Edistr} one gets the kinetic energy,
\begin{align}\label{Ekinosc}
  \langle E_k \rangle \sim &
  \left(
    \int \mathrm{d}^3\mathbf{r} |n_1|
  \right)^{-1}
  \notag
  \\
  & \times
  \int \mathrm{d}^3\mathbf{r} |n_1| \frac{m |\mathbf{v}_1|^2}{2}
  \notag
  \\
  & \approx
  0.08 \frac{e^2 A^2 \sigma^2}{m \omega_p^2},
\end{align}
where we take into account the relation between the main harmonic
amplitudes of the electron gas presented in Eq.~\eqref{vEn}. Note
that in Eq.~\eqref{Ekinosc} we take into account only the part of
the kinetic energy due to the oscillatory motion rather than the
total energy which also includes the thermal contribution.

To analyze the behaviour of the effective interaction first we
mention that one can explicitly calculate the common integral in
Eqs.~\eqref{toten} and~\eqref{Mfinal},
\begin{align}\label{aux}
  \int_0^\infty & \mathrm{d}r \ r U(r) \sin kr
  \notag
  \\
  & =
  - \frac{\alpha \pi^2 e^2}{r_0} \mathrm{e}^{-k r_0}
  \notag
  \\
  & \times
  \int_0^\infty \mathrm{d}r \ r (\sin kr - kr \cos kr) E_1(r)
  \notag
  \\
  & =
  \frac{\alpha e A \sigma^5 k^3}{4r_0}
  \sqrt{\frac{\pi}{2}} \mathrm{e}^{-k r_0}
  \exp
  \left(
    -\frac{k^2\sigma^2}{2}
  \right),
\end{align}
where we use Eqs.~\eqref{vEn}, \eqref{Edistr}, and~\eqref{Upot}.
Then we rewrite Eq.~\eqref{2ordme} as $V_{k k' q}/\langle E_k
\rangle = W/(R-1)$, where we introduce the new functions,
\begin{align}\label{RW}
  R = &
  \left(
    \frac{\Delta E}{\hbar \omega_q}
  \right)^2
  \approx \frac{0.5 \times 10^{-2}}{\xi^4}
  \notag
  \\
  & \times
  \left(
    \frac{\sigma}{10^{-7}\thinspace\text{cm}}
  \right)^{-4}
  f_R(k\sigma),
  \notag
  \\
  W = & \frac{2 |M_{k,k-q}|^2}{\hbar \omega_q \langle E_k \rangle}
  \approx \frac{3.1 \times 10^{-2}}{\xi^2}
  \notag
  \\
  & \times
  \left(
    \frac{n_a}{10^{21}\thinspace\text{cm}^{-3}}
  \right)^2
    \left(
    \frac{\sigma}{10^{-7}\thinspace\text{cm}}
  \right)^{-3}
  \notag
  \\
  & \times
  f_W(k\sigma).
\end{align}
Here $f_R(x) = x^{16} \mathrm{e}^{-2x^2}$ and $f_W(x) = x^{11}
\mathrm{e}^{-x^2}$. To derive Eq.~\eqref{RW} we use
Eq.~\eqref{aux} and the following plasma parameters: $\alpha \sim
10^{-24}\thinspace\text{cm}^3$, $r_0 \sim
10^{-8}\thinspace\text{cm}$, $\sigma_s \sim
10^{-15}\thinspace\text{cm}^2$, $M_a \sim
10^{-24}\thinspace\text{g}$, and $c_a \sim
10^5\thinspace\text{cm/s}$. The effective interaction turns out to
be attractive when $R < 1$. The function $W$ is the ``magnitude''
of the interaction expressed in terms of the mean kinetic energy
of oscillating electrons, cf. Eq.~\eqref{Ekinosc}.

\begin{figure}
  \centering
  \includegraphics[scale=.4]{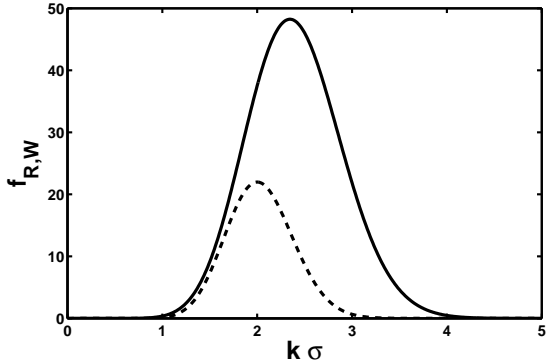}
  \caption{\label{fRfW}
  The functions $f_R$ (dashed line) and $f_W$ (solid line) versus
  $k\sigma$.}
\end{figure}

In Fig.~\ref{fRfW}
we show the behavior of the functions $f_{R,W}$ versus the
dimensionless variable $k\sigma = \omega \sigma / c_a$.  As we can
see these functions have the maxima due to the enhanced emissivity
of acoustic waves with the wave length of the order of the size of
a plasmoid, $2 \pi /k \sim \sigma$.
Using Eq.~\eqref{RW} and Fig.~\ref{fRfW} we get the lower bound on
the width of a plasmoid at which the effective attraction between
electrons takes place, $\sigma > 0.3 \times 10^{-7} \thinspace
\text{cm} \times \xi^{-1}$. To obtain this constraint we account
for the maximum of the function $(f_R)_\mathrm{max} \approx 22$.

In our analysis we take into account only the interaction of
oscillating electrons with neutral atoms. It is however known that
oscillations of plasma are possible if the total number of
collisions of background electrons per unit time $\nu$ is less
than the oscillations frequency. Note that one should account for
collisions only inside the plasmoid. Thus we get for this
quantity, $\nu = v_\mathrm{T} \sigma_s n_a V_\mathrm{eff} n_0$,
where $v_\mathrm{T} = \sqrt{T/m}$ is the thermal electron
velocity, $V_\mathrm{eff} = (4 \pi/3) R_\mathrm{eff}^3 = 17
\sigma^3$ is the effective plasmoid volume, and $R_\mathrm{eff} =
1.6 \sigma$ is the effective plasmoid radius calculated as
$R_\mathrm{eff}^2 = \textstyle{\int E^2 r^2
\mathrm{d}^3\mathbf{r}} / \textstyle{\int E^2
\mathrm{d}^3\mathbf{r}}$, cf. Eq.~\eqref{Edistr}.

Taking $T = 10^3\thinspace\text{K}$ and using Eq.~\eqref{RW} at
the maximal value of the function $W$, corresponding to the
strongest effective attraction, we get that plasma oscillations
are possible (i.e. the condition $\nu < \omega$ is satisfied) if
$\sigma < 1.4 \times 10^{-6} \thinspace \text{cm} \times
\xi^{2/7}$. Combining this result with the previously obtained
lower bound for $\sigma$, one gets that the width of a plasma
structure should be in the range, $0.3 \times 10^{-7} \thinspace
\text{cm} \times \xi^{-1} < \sigma < 1.4 \times 10^{-6} \thinspace
\text{cm} \times \xi^{2/7}$. Therefore, for the effective
attraction between oscillating electrons to take place, the
frequency of oscillations should lie in the following interval:
$0.05 \omega_p < \omega < \omega_p$. Note that for the chosen
parameters the Debye length for electrons $\lambda_\mathrm{D} =
v_\mathrm{T}/\omega_p$ is of the order of $R_\mathrm{eff}$.

\section{Applications\label{APPL}}

In Sec.~\ref{PAF} we showed that the exchange of a virtual
acoustic wave between two electrons participating in spherically
symmetric oscillations may result in their effective attraction.
The typical size of the plasma structure when the effective
attraction becomes important, i.e. when it is comparable with the
mean kinetic energy of oscillating particles, is quite small $\sim
10^{-6}\thinspace\text{cm}$. Although in the present work we do
not analyze the physical processes which underlie the plasmoid
existence, such a small size is the indication to the quantum
nature of this plasma structure. For example, quantum plasmoids of
the similar size were described by~\citet{DvoDvo07} on the basis
of the solution of the nonlinear Schr\"{o}dinger equation.

If the effective attraction between oscillating electrons is
sufficiently strong, we may suggest that these particles can form
a bound state. Previously~\citet{NamAka85,NamVlaShu95} suggested
that bound states of charged particles in plasma can be formed due
to the exchange of ion acoustic waves. Another mechanism based on
the magnetic interaction was proposed by~\citet{Mei84}. Note that
in our case the presence of a strong nonlinear effect is required
to significantly diminish the frequency of oscillations, $\omega <
\omega_p$. We suggest that the phenomenon of the bound states
formation of electrons in plasma can be implemented inside a
stable atmospheric plasmoid~\citep{Ste99}, called a ball lightning
(BL). These kind of plasma structures appears during a
thunderstorm and have the lifetime up to several minutes. Despite
a great variety of models of BL summarized
by~\citet[pp.~270--296]{BycNikDij10}, these objects are likely to
be plasma based phenomena.

To describe the long lifetime of a natural plasmoid,~\citet{Dij80}
suggested that BL is a spherical vortex composed of a dense
superconducting plasma. Recently~\citet{Zel08} revisited the idea
that the plasma superconductivity may be implemented in BL.
\citet{Zel08}~suggested that a natural plasmoid can have the
positively charged kernel and the superconducting electron
envelop. Nevertheless those models were based on the
phenomenological assumption of the plasma superconductivity
without pointing out a physical mechanism which underlies it.

As we mentioned in Sec.~\ref{EIG} the incoherent emission of
acoustic waves results in the energy dissipation in the plasmoid.
On the contrary, the coherent exchange by an acoustic wave leads
to the attractive interaction between electrons. As we showed in
Sec.~\ref{PAF}, this attraction can result in the formation of
bound states, analogous to Cooper pairs in metals~\citep{Mad80}.
Thus one may expect that the electron component of plasma will
experience the phase transition followed by the formation of the
condensate of electrons. The collective oscillations of the
electrons condensate can be responsible for reducing the
resistance of plasma and thus can provide the observed lifetime of
a natural plasma structure. Although this hypothesis requires the
additional analysis which would carefully account for all thermal
effects, the estimates given in Sec.~\ref{PAF} show that the
described process is quite possible.

For the existence of plasma oscillations the number of collisions
of oscillating particles per unit time should be much less than
the oscillations frequency, $\nu \ll \omega$. Otherwise
oscillations will decay. In Sec.~\ref{PAF} we checked that the
weaker inequality, $\nu < \omega$, is satisfied. However, if a
plasma is in the reduced resistance state and supposing that
rather strong nonlinearity is present, i.e. a significant fraction
of electrons participates in oscillations, we may expect that the
condition $\nu \ll \omega$ is satisfied as well. In Sec.~\ref{PAF}
we also obtained that for the chosen parameters the Debye length
is of the order of the plasmoid size. If $R_\mathrm{eff} \ll
\lambda_\mathrm{D}$, oscillating electrons sometimes can pass
through the plasma structure that results in the energy
dissipation~\citep{Gol84}. Although we do not violate the
condition $R_\mathrm{eff} \gtrsim \lambda_\mathrm{D}$, the
excessive energy dissipation can be avoided in the reduced
resistance plasma state.

In the present work we considered a plasmoid with a very small
core $\sim 10^{-6}\thinspace\text{cm}$, whereas according to
observations the visible diameter of a natural plasma structure is
of the order of several centimeters~\citep{BycNikDij10}.
Nevertheless, if one analyzes the available photographs of this
phenomenon published by~\citet[pp.~129--161]{Bur07,Ste99}, one can
conclude that, while moving in the atmosphere, a plasmoid leaves a
trace which is much smaller than its visible size. It can be
possible if a small and hot kernel exists inside a plasmoid. This
core ionizes the air and produces the observed trace. The visible
dimensions of a plasmoid are likely to be attributed to auxiliary
effects.

One more indication that the actual size of a natural plasma
structure is much smaller than the visible one, is in the fact
that sometimes it can pass through tiny holes and cracks, with the
structure of the plasmoid being
unchanged~\citep[pp.~203--246]{BycNikDij10}. In many cases the
materials which a plasmoid passes through are not damaged either.
The most natural explanation of this unusual behavior is the
suggestion of the small scale internal structure of the object.
Although it is not directly related to our description of natural
plasmoids, we should mention that recently~\citet{Mul10} discussed
the model of BL in which the visible size of the object is bigger
than its core having the radius $\sim 0.1-2\thinspace\text{cm}$.

Another evidence of the small-sized core of long-lived plasma
structures can be obtained from the laboratory
experiments~\citep[pp.~263--265]{BycNikDij10}. Moreover
recently~\citet{Kli10} reported that plasmoids with a nano-scale
kernel were generated in the studies of high frequency discharges
in dusty plasmas. The experiments with silicon discharges carried
out by~\citet{Abr02,Laz06,DikJer06,Pai07,Mit08}, where glowing
structures resembling natural plasmoids were generated, should be
also mentioned. Although the interpretation of the results of
those experiments involves another plasmoid model, the generated
objects have nano-sized cores as well.

The reports of the energy content of natural plasma structures are
rather different: along the objects with relatively small internal
energy, the plasmoids possessing huge energy were
observed~\citep[pp.~203--246]{BycNikDij10}. The analysis of the
present work is applicable for a low energy plasma structure which
does not seem to have an internal energy source. That is why in
Sec.~\ref{PAF}, while making numerical estimates, we took the
temperature of electrons $\sim 10^3\thinspace\text{K}$. We remind
that the maximal electron temperature is $\sim
10^5\thinspace\text{K}$. Note that the existence of low
temperature plasmoids is not excluded by
observations~\citep[pp.~203--246]{BycNikDij10}. It is worth
mentioning that, if the plasma of such a low energy BL is in the
superconducting state, one can avoid the energy losses, plasma
recombination, and provide the plasma structure stability.

We showed that the model of a nano-sized plasmoid is able to
describe some of the observed properties of BL. Nevertheless we
may also suggest that in a realistic natural plasma structure
there could be multiple tiny kernels where intense electron
oscillations happen. Note that the analogous model of a composite
BL was discussed by~\citet{Nik06}. The separate oscillatory
centers can be held together by the attractive quantum exchange
forces~\citep{KulRum91}, which are relevant in our situation since
the predicted size of a single core is tiny, $\sim
10^{-6}\thinspace\text{cm}$. However the detailed description of
the coagulation process using the results of~\citet{KulRum91}
requires additional special analysis.

The analysis of the observed characteristics of natural plasmoids
summarized by~\citet[pp.~203--246]{BycNikDij10} shows that they
are likely to be the phenomena of different origin. Therefore a
unique model which would explain all the observations does not
seem to exist. For instance, the plasmoid model described in the
present work does not explain the electromagnetic action of BL,
like the generation of strong electric currents and the emission
of radio-waves. Nevertheless in frames our approach we could
naturally explain some of the BL properties
which are hard to account for in alternative models of natural
plasmoids.

At the end of this section we may say a few words how an
atmospheric plasmoid based on spherically symmetric oscillations
of electrons appears in natural conditions. As we revealed in
Sec.~\ref{PAF}, for the existence of a plasmoid the frequency of
plasma oscillations should be comparable with $\omega_p$ which is
in the GHz region. It is very difficult to generate such a high
frequency during a thunderstorm since, e.g., a linear lightning is
a quite low frequency phenomenon~\citep{RakUma06}. Thus the
possible BL generation scenario looks as follows. Suppose that
during a linear lightning stroke a natural capacitor with a small
capacity $\sim 10\thinspace\text{pF}$ is charged up to a very high
voltage. Then, if this capacitor is discharged on a thin point, a
GHz electromagnetic oscillation can be created, provided the
discharge channel inductance is $~\sim 0.1\thinspace\mu\text{H}$.
However many other factors, like shape of the point, air humidity
etc, should be properly combined for the successful BL generation.

\section{Conclusion\label{CONCL}}

In this work we have discussed the effective interaction between
electrons in a low temperature plasma due to the exchange of
virtual acoustic waves. The electron temperature of such a plasma
should be low enough to allow the existence of both electrons,
ions, and neutral atoms. In Sec.~\ref{EIG} we have derived the
expression for the potential of this effective
interaction~\eqref{2ordme} and considered a particular case of the
effective interaction between electrons participating in
spherically symmetric oscillations. We supposed that charged
particles in plasma perform nonlinear oscillations and form a
stable plasma structure. However in our analysis we just used the
commonly adopted \emph{ansatz} for the electric field
distribution~\eqref{Edistr} without going into details what kind
of physical processes provides the plasmoid stability.

Then, in Sec.~\ref{PAF}, we have evaluated the parameters of the
effective interaction. It has been established that under certain
conditions the interaction is attractive and its strength can be
comparable with the mean kinetic energy of oscillating electrons.
For this situation to happen, the typical size of a plasmoid
should be quite small, $\sim 10^{-6}\thinspace\text{cm}$. It is
the indication that one should use the concept of quantum plasmas
to describe the stability of the plasma structure.

Note that analogous effective interaction will also appear between
ions and neutral atoms as well as between ions and electrons since
Eq.~\eqref{2ordme} is of the universal type. However, in the
former case the frequency of the ion oscillations is significantly
less than $\omega_p$ and the interaction will be always repulsive.
In the latter case we should recall that electrons and ions
interact mainly electromagnetically and the small contribution due
to the collisions will be negligible.

Finally, in Sec.~\ref{APPL}, we have discussed the possible
application of the obtained results to the description of stable
natural plasma structures~\citep{Ste99,BycNikDij10}. According to
the model of~\citet{DvoDvo07,Dvo10,Dvo11} these objects can be
implemented as spherically symmetric plasma oscillations. Note
that analogous idea was also discussed by~\citet{Shm03}. To
explain the long lifetime (up to several minutes) of a natural
plasmoid we put forward a hypothesis that plasma could be in a
reduced resistance state. The mechanism underlying the existence
of such a state could be the exchange of virtual acoustic waves
between oscillating electrons, described in Secs.~\ref{EIG}
and~\ref{PAF}. We have also considered how the characteristics of
a plasma structure, predicted in frames of our model, conform to
the observed properties of natural plasmoids.

\section*{Acknowledgments}

This work has been supported by CONICYT (Chile) through Programa
Bicentenario PSD-91-2006. The author is thankful to FAPESP
(Brazil) for a grant as well as to S.~I.~Dvornikov, A.~I.~Nikitin,
and V.~B.~Semikoz for helpful discussions.

\balance

\end{document}